\documentclass{iopjournal}

\usepackage{graphicx}
\usepackage{bm}
\usepackage{ragged2e}
\usepackage{lipsum}
\usepackage{amsmath}
\usepackage{amsthm}
\usepackage{mathrsfs}
\usepackage{amsfonts}
\usepackage{amssymb}
\usepackage{textcomp}
\usepackage{appendix}
\usepackage{cite}

\begin{document}

\articletype{Article type} 

\title{Practical implementation of arbitrary nonlocal controlled-unitary gate via indefinite causal order}

\author{Wen-Qiang Liu$^{1,*}$\orcid{0000-0001-9804-2716}, Zi-Han Zheng$^1$\orcid{0009-0008-4414-4213}, Zhang-Qi Yin$^2$\orcid{0000-0001-6260-3083} and Hai-Rui Wei$^{3,*}$\orcid{0000-0001-7459-4161}}

\affil{$^1$Department of Mathematics and Physics, Shijiazhuang Tiedao University, Shijiazhuang 050043, China}

\affil{$^2$Center for Quantum Technology Research and Key Laboratory of Advanced Optoelectronic Quantum Architecture and Measurements (MOE), School of Physics, Beijing Institute of Technology, Beijing 100081, China}

\affil{$^3$School of Mathematics and Physics, University of Science and Technology Beijing, Beijing 100083, China}

\affil{$^*$Author to whom any correspondence should be addressed.}

\email{wqliu@stdu.edu.cn and hrwei@ustb.edu.cn}


\begin{abstract} \justifying
Quantum gate teleportation enables the implementation of nonlocal quantum operations without direct interactions between distant nodes. 
We propose an efficient protocol for implementing arbitrary controlled-unitary (CU) gates acting on two spatially separated parties via indefinite causal order (ICO).
By establishing a maximally entanglement  between two remote nodes and coherently superposing orders of single-qubit gates,
our protocol circumvents the drawback of complex local two-qubit operations.
This ICO-based approach enables full programmability of CU gates by adjusting the inherent single-qubit operations, offering advantages over conventional fixed causal-order methods in terms of reduced circuit complexity and improved experimental flexibility.
Furthermore, we develop an optical construction to implement the polarization CU gate using a stable and reciprocal Sagnac interferometer.
Our work establishes a practical framework for scalable distributed quantum computation with flexible operations.
\end{abstract}

\section{Introduction}  \label{sec1}

\justifying

Quantum computing promises significant advantages in addressing certain problems that are intractable for classical counterparts \cite{nielsen2010quantum}, particularly in the noisy intermediate-scale quantum era \cite{arute2019quantum,zhong2020quantum,deng2023gaussian,liu2024efficiently}.
In recent years, quantum computing has gained significant attention, and accompanied by notable advancements in local qubit arrays \cite{knill2001scheme,o2007optical,liu2020low,cong2022hardware,Litao2024,20255,2025DU}.
However, nonlocal quantum computing remains an area that is in need of further exploration.
Building a large-scale quantum computer also poses the key technical challenge, that because precise qubit manipulation and effective coupling are hindered by residual crosstalk and spatial constraints within a single device \cite{ladd2010quantum}.
To exceed these limitations, distributed quantum computing (DQC) \cite{lim2005repeat,jiang2007distributed,oh2023distributed} has emerged as an effective strategy for constructing robust and scalable quantum systems within a large-scale quantum network.
In a DQC architecture, the composite system is modularized into optimized subsystems that operate  locally, and the subsystems are interconnected via quantum state transfer. Notably, such modularization can suppress crosstalk and redundant operations across the network, although noise may still propagate between modules  \cite{murali2020software,avron2021quantum,akahoshi2024partially}.


Quantum gate teleportation (QGT) \cite{eisert2000optimal,bartlett2003quantum,hu2023progress,carrera2024combining} is a core technology for DQC, that of enabling the nonlocal quantum gates across network nodes to overcome the challenges of realizing the long-range interactions in quantum information processing tasks \cite{gottesman1999demonstrating}.
With the help of quantum teleportation \cite{pirandola2015advances}, QGT can be realized by employing a pre-shared entanglement between two remote nodes, local two-qubit gates, together with local operations and classical communications \cite{huang2004experimental,gao2010teleportation,chou2018deterministic,wan2019quantum,daiss2021quantum,liu2024nonlocal,Feng2024,main2025distributed,qiu2025deterministic}.
The teleportation of a controlled-NOT (CNOT) gate was first experimentally demonstrated in photonic systems \cite{huang2004experimental,gao2010teleportation}.
Later, CNOT gate teleportation has been experimentally reported across diverse physical platforms, including superconducting qubits \cite{chou2018deterministic},  trapped ions \cite{wan2019quantum}, and atom-cavity systems \cite{daiss2021quantum}.
More recently, the photonic CNOT gate teleportation over 7.0 $km$ \cite{liu2024nonlocal}, chip-to-chip photonic CNOT gate teleportation \cite{Feng2024}, and CNOT gate teleportation in other systems \cite{main2025distributed,qiu2025deterministic}, have also been experimentally demonstrated.

The current existing architectures of CNOT gate teleportations universally adhere to the fixed causal-order framework established, where local CNOT operations at each network node are indispensable \cite{huang2004experimental,gao2010teleportation,chou2018deterministic,daiss2021quantum,wan2019quantum,liu2024nonlocal,Feng2024,main2025distributed}.
 Recently, indefinite causal order (ICO) has emerged as a novel quantum resource \cite{brukner2015bounding,zych2019bell,taddei2019quantum,chiribella2021indefinite,liu2023quantum,rozema2024experimental,salzger2025mapping} and has been shown to enable nonlocal quantum operations \cite{ghosal2023quantum,liu2025quantum,liu2025deterministic}.
In conventional remote quantum circuit model, teleportation of arbitrary two-qubit controlled-unitary (CU) gate can be achieved by replacing the local CNOT gate placed at the second node with a local CU gate \cite{eisert2000optimal,liu2024nonlocal}. Recently, the CNOT gate teleportation based on the ICO approach has been proposed \cite{liu2025quantum}. A direct implementation of CU gate teleportation based on the standard decomposition is considerably more challenging than that of a CNOT gate, since the realization of a general CU gate requires two CNOT gates and three fundamental single-qubit operations \cite{barenco1995elementary}.
To date, no simpler or more cost-effective method for CU gate teleportation has been identified, and existing approaches remain far from experimental feasibility.

In this paper, we propose a deterministic general CU gate teleportation protocol within an ICO quantum circuit.
The protocol is implemented by a pre-sharing  maximally entangled state between two distant nodes, together with the superposition of single-qubit gate operations and classical communications.
It achieves the minimal communication cost, requiring only one entanglement bit (ebit) and two classical bits (cbits) \cite{jiang2007distributed,eisert2000optimal}. 
Our protocol bypasses the need of complex local CNOT and CU gates in traditional protocols \cite{eisert2000optimal, liu2024nonlocal} by governing simpler single-qubit gates.
The protocol reduces the resource cost from two ebits, four cbits, and four quantum switches to  one ebit, two cbits, and two quantum switches and enhances experimental feasibility.
Furthermore, we develop a compact optical scheme for teleporting the CU gate encoded in the polarization degree of freedom (DOF) by using the reciprocal Sagnac-type interferometers.

\section{The protocol of an arbitrary CU gate teleportation} \label{Sec2}

QGT enables two distant nodes to collectively implement a nonlocal quantum gate without requiring any direct interaction between them.
Consider two parties, Alice and Bob, who are spatially separated. Alice holds qubit $A$, which is initially prepared in the state
\begin{eqnarray}  \label{eq1}
|\varphi\rangle_A =\alpha_1|0\rangle_A+\beta_1|1\rangle_A.
\end{eqnarray}
Bob holds another qubit $B$,  which is prepared in the state
\begin{eqnarray}  \label{eq2}
|\varphi\rangle_B =\alpha_2|0\rangle_B+\beta_2|1\rangle_B.
\end{eqnarray}
The arbitrary complex coefficients $\alpha_i$  and $\beta_i$ ($i=1, 2$) satisfy the normalization condition.
Alice and Bob wish to perform an arbitrary nonlocal CU gate on the joint state $|\varphi\rangle_A\otimes|\varphi\rangle_B$, where qubit $A$ acts as the control qubit and qubit $B$ acts as the target qubit.
An arbitrary CU gate is defined as \cite{nielsen2010quantum}
\begin{eqnarray}  \label{eq3}
\text{CU} = |0\rangle_A\langle0|\otimes \mathbb{I}_B+|1\rangle_A\langle1|\otimes U_B(\alpha, \theta, \textbf{n}),
\end{eqnarray}
where $\mathbb{I}_B$ represents  the identity operator acting on qubit $B$,
and $U_B(\alpha, \theta, \textbf{n})=\text{exp}\big[\texttt{i}\big(\alpha \mathbb{I}+\theta(\textbf{n}\cdot \bm{\sigma})\big)\big]$ is an arbitrary single-qubit unitary gate acting on qubit $B$.
Here, $\alpha$ and $\theta$ are real parameters, $\mathbf{n} = (n_x, n_y, n_z)$ is a real unit vector in three dimensions, and $\bm{\sigma} = (X, Y, Z)$ represents the Pauli vector.
Specifically, if the control qubit $A$ is in the state $|1\rangle$, the gate $U_B$ is applied to the target qubit $B$; otherwise, $B$ remains unchanged.

\begin{figure} 
\begin{center}
\includegraphics[width=13.5 cm,angle=0]{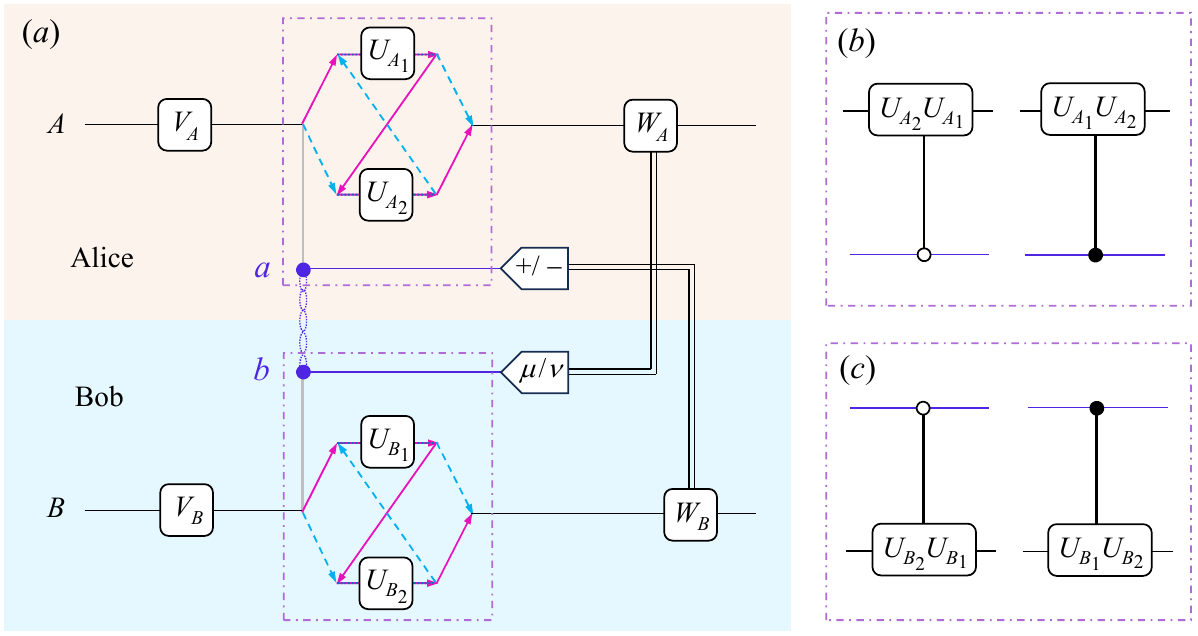}
\caption{\justifying
(a) Schematic diagram for teleporting an arbitrary CU gate acting on two spatially separated qubits, $A$ and $B$.
A maximally entangled state $|\varphi\rangle_{ab} = \frac{1}{\sqrt{2}}(|0\rangle_a|0\rangle_b + \texttt{i}|1\rangle_a|1\rangle_b)$ is shared between Alice and Bob. The ancillary qubits $a$ and $b$ coherently control the order of single-qubit gate operations via quantum switches, shown as purple boxes.
If qubit $a$ (or $b$) is in the state $|0\rangle_a$ (or $|0\rangle_b$), the gate $U_{A_1}$ (or $U_{B_1}$) is applied before $U_{A_2}$ (or $U_{B_2}$), i.e., the pink solid circuit.
If qubit $a$ (or $b$) is in the state $|1\rangle_a$ (or $|1\rangle_b$), the gate $U_{A_2}$ (or $U_{B_2}$) is applied before $U_{A_1}$ (or $U_{B_1}$), i.e., the blue dashed circuit. By measuring the ancillary qubits in appropriate bases and applying corresponding feed-forward operations, the teleportation of the CU gate is completed. The single-qubit gates $U_{A_1}=R_{z}(\frac{\pi}{2})$, $U_{A_2}=X$, $U_{B_1}=R_\textbf{n}(\frac{\pi}{2})$, and $U_{B_2}=\mathbf{n}^\perp\cdot \bm{\sigma}$. The quantum switches at Alice’s and Bob’s nodes are schematically illustrated in panels (b) and (c), where the symbols ``$\circ$'' and ``$\bullet$'' denote control qubits in the states $|0\rangle$ and $|1\rangle$, respectively.} \label{nonlocalCU}
\end{center}
\end{figure}

Figure~\ref{nonlocalCU} illustrates the schematic procedure for teleporting a CU gate via ICOs.
Let us now introduce the step-by-step operation of the protocol.

Firstly, a maximally entangled state
\begin{eqnarray}
\begin{split}             \label{eq4}
|\varphi\rangle_{ab}=\frac{1}{\sqrt{2}}(|0\rangle_a|0\rangle_b+\texttt{i}|1\rangle_a|1\rangle_b),
\end{split}
\end{eqnarray}
served as an ancillary quantum state is shared with Alice and Bob.
Qubit $a$ is sent to Alice and qubit $b$ to Bob. Consequently, Alice holds qubits $A$ and $a$, while Bob possesses qubits $B$ and $b$.
The initial state of the whole system is thus given by
\begin{eqnarray}
\begin{split}             \label{eq5}
|\Phi\rangle_{0}=|\varphi\rangle_{ab}\otimes|\varphi\rangle_A\otimes|\varphi\rangle_B.
\end{split}
\end{eqnarray}
Next, Alice applies a single-qubit gate $V_A = X$ to her qubit $A$, and Bob applies a gate $V_B = \mathbf{n}^\perp \cdot \bm{\sigma}$ to his qubit $B$.
These local operations, $V_A$ and $V_B$,  transform the global state $|\Phi\rangle_{0}$ into
\begin{eqnarray}
\begin{split}             \label{eq6}
|\Phi\rangle_{1}=\frac{1}{\sqrt{2}}(|0\rangle_a|0\rangle_b+\texttt{i}|1\rangle_a|1\rangle_b)\otimes |\varphi'\rangle_A\otimes |\varphi'\rangle_B.
\end{split}
\end{eqnarray}
Here $|\varphi'\rangle_A=V_A|\varphi\rangle_A$ and $|\varphi'\rangle_B=V_B|\varphi\rangle_B$.
$\mathbf{n}^\perp=(n_x', n_y', n_z')$ is a unit vector orthogonal to the vector $\mathbf{n}$, i.e., $\mathbf{n} \cdot \mathbf{n}^\perp = 0$.



Secondly, qubits $A$ and $B$ undergo local quantum switch operations \cite{chiribella2013quantum,goswami2018indefinite,liu2023experimentally,rozema2024experimental}, each coherently controlled by the corresponding auxiliary qubits, $a$ and $b$, respectively. At Alice's node, the switch applies local single-qubit gates $U_{A_1}$ and $U_{A_2}$ on qubit $A$ in an order determined by the state of $a$. That is, if qubit $a$ is in the state $|0\rangle_a$, the gate order is $U_{A_2}U_{A_1}$ (pink solid circuit); if qubit $a$ is in the state $|1\rangle_a$, the order is $U_{A_1}U_{A_2}$ (blue dashed circuit). A similar control mechanism operates at Bob's node, where  additional qubit $b$ manipulates the order of $U_{B_1}$ and $U_{B_2}$ on qubit $B$.   As a result, the joint action of the two quantum switches transforms $|\Phi\rangle_{1}$ into
\begin{eqnarray}
\begin{split}             \label{eq7}
|\Phi\rangle_{2}=&\frac{1}{\sqrt{2}}(|0\rangle_a|0\rangle_b\otimes U_{A_2}U_{A_1}|\varphi'\rangle_A\otimes U_{B_2}U_{B_1}|\varphi'\rangle_B \\&
+\texttt{i}|1\rangle_a|1\rangle_b\otimes U_{A_1}U_{A_2}|\varphi'\rangle_A \otimes U_{B_1}U_{B_2}|\varphi'\rangle_B).
\end{split}
\end{eqnarray}
In order to implement the desired nonlocal CU gate, we define these single-qubit gates as follows
\begin{eqnarray}
\begin{split}             \label{eq11}
U_{A_1} =R_z(\frac{\pi}{2})=\left(\begin{array}{cc}
          e^{-\texttt{i}\frac{\pi}{4}}    &           0                 \\
                 0                        &  e^{\texttt{i}\frac{\pi}{4}}\\
\end{array}\right),\\
\end{split}
\end{eqnarray}
\begin{eqnarray}
\begin{split}             \label{eq11.1}
U_{A_2}=X=\left(\begin{array}{cc}
           0  &  1  \\
           1  &  0  \\
\end{array}\right), \\
\end{split}
\end{eqnarray}
\begin{eqnarray}
\begin{split}             \label{eq11.2}
U_{B_1}=R_\textbf{n}(\frac{\pi}{2})=\frac{1}{\sqrt{2}} \left(\begin{array}{cc}
             1-\texttt{i} n_z            &  -\texttt{i}n_x-n_y\\
            -\texttt{i} n_x+n_y  &   1+\texttt{i} n_z     \\
\end{array}\right),\\
\end{split}
\end{eqnarray}
and
\begin{eqnarray}
\begin{split}             \label{eq11.3}
U_{B_2}=\mathbf{n}^\perp\cdot \bm{\sigma}=\left(\begin{array}{cc}
           n_z'  &  n_x'-\texttt{i} n_y'  \\
           n_x'+\texttt{i} n_y'  &  -n_z'  \\
\end{array}\right).
\end{split}
\end{eqnarray}

Finally, the ancillary qubits $a$ and $b$ are measured in the bases $\{|\pm\rangle\}$ and $\{|\mu(\theta)\rangle, \nu(\theta)\rangle\}$, respectively. Here $|\pm\rangle=\frac{1}{\sqrt{2}}(|0\rangle\pm|1\rangle)$ and the basis $\{|\mu(\theta)\rangle, \nu(\theta)\rangle\}$ is given by
\begin{eqnarray}
\begin{split}             \label{eq8}
|\mu(\theta)\rangle=\cos\big(\frac{\theta}{2}\big)|0\rangle+\sin\big(\frac{\theta}{2}\big)|1\rangle, \quad |\nu(\theta)\rangle=\sin\big(\frac{\theta}{2}\big)|0\rangle-\cos\big(\frac{\theta}{2}\big)|1\rangle.
\end{split}
\end{eqnarray}
If Alice obtains the measurement outcome $|+\rangle_a$, she sends a classical bit to inform Bob of her result. Bob then measures his qubit $b$ in the basis $\{|\mu(\theta)\rangle, |\nu(\theta)\rangle\}$.
Alternatively, if Alice obtains the outcome $|-\rangle_a$, she communicates this result to Bob, and then Bob measures his qubit $b$ in the basis $\{|\mu(\pi-\theta)\rangle, |\nu(\pi-\theta)\rangle\}$. This adaptivity preserves the quantum superposition of gate orders and does not introduce the causal asymmetry.

For the coincidence measurement outcomes $|+\rangle_a|\mu(\theta)\rangle_b$ or $|-\rangle_a|\nu(\pi-\theta)\rangle_b$, the state $|\Phi\rangle_{2}$ is projected onto
\begin{eqnarray}
\begin{split}             \label{eq9}
|\Phi\rangle_{3}=\big[\cos\big(\frac{\theta}{2}\big) U_{A_2}U_{A_1}\otimes U_{B_2}U_{B_1} +\texttt{i} \sin\big(\frac{\theta}{2}\big)  U_{A_1}U_{A_2}\otimes U_{B_1}U_{B_2}\big]|\varphi'\rangle_A|\varphi'\rangle_B,
\end{split}
\end{eqnarray}
with a probability of $\frac{1}{2}$.
On the other hand, for the coincidence  measurement outcomes $|+\rangle_a|\nu(\theta)\rangle_b$ or $|-\rangle_a|\mu(\pi-\theta)\rangle_b$, the state $|\Phi\rangle_{2}$ is projected onto
\begin{eqnarray}
\begin{split}             \label{eq10}
|\Phi'\rangle_{3}=\big[\sin\big(\frac{\theta}{2}\big) U_{A_2}U_{A_1}\otimes U_{B_2}U_{B_1}  -\texttt{i} \cos\big(\frac{\theta}{2}\big)  U_{A_1}U_{A_2}\otimes U_{B_1}U_{B_2}\big]|\varphi'\rangle_A|\varphi'\rangle_B,
\end{split}
\end{eqnarray}
with a probability of $\frac{1}{2}$.
By substituting equations~\eqref{eq11}-\eqref{eq11.3} into equations~\eqref{eq9} and \eqref{eq10}, we obtain an effective gate operation that is equivalent to a nonlocal CU gate acting on the initial product state $|\varphi\rangle_A \otimes |\varphi\rangle_B$.
The detailed proof is provided in the appendix.
%
%

The relationships between the feed-forward operations (performed on qubits $A$ and $B$ to complete the protocol) and the corresponding measurement outcomes are summarized in table~\ref{table1}.
More specifically,
if the measurement outcomes are $|+\rangle_a|\mu(\theta)\rangle_b$ or $|-\rangle_a|\nu(\pi-\theta)\rangle_b$, Alice applies unitary operation $W_{A}=R_z(\alpha+\frac{\pi}{2})$ on qubit $A$, and Bob applies $W_B = R_{\mathbf{n}}(\frac{\pi}{2} - \theta)$ on qubit $B$ to implement the CU gate, up to a global phase $e^{\texttt{i}\frac{\alpha}{2}}$.
If the measurement outcomes are $|+\rangle_a|\nu(\theta)\rangle_b$ or $|-\rangle_a|\mu(\pi-\theta)\rangle_b$, Alice applies $W_{A}=R_z(\alpha-\frac{\pi}{2})$ on qubit $A$, and Bob applies $W_{B}=R_\mathbf{n}(-\frac{\pi}{2}-\theta)$ on qubit $B$ to implement the CU gate, up to a global phase $\texttt{i}e^{\texttt{i}\frac{\alpha}{2}}$.
Here
\begin{eqnarray}             \label{eq13}
R_\mathbf{n}(\theta)=\cos\big(\frac{\theta}{2}\big)\mathbb{I}-\texttt{i}\sin\big(\frac{\theta}{2}\big)(\textbf{n}\cdot \bm{\sigma}),
\end{eqnarray}
represents a single-qubit rotation by an angle $\theta$ around the \textbf{n}-axis.
In particular, $R_z(\theta)$ denotes a rotation by angle $\theta$ around the $z$-axis, corresponding to the choice $\mathbf{n}_z = (0, 0, 1)$.


\begin{table}  \renewcommand{\arraystretch}{1.3}  
\centering \caption{\justifying
Correspondence between the measurement outcomes on qubits $a$ and $b$ and the required feed-forward operations on qubits $A$ and $B$ for implementing an arbitrary CU gate. The final output state differs from the ideal CU operation only by a  trivial global phase.}
\begin{tabular}{ccccccc}
\hline

\multicolumn {2}{c}{Measurement}  & \;\,\,\, & \multicolumn {2}{c}{Feed-forward} & \;\,\,\,  & Global  \\ \cline{1-2}  \cline{4-5}

   Qubit $a$   &      Qubit $b$      & \quad &   $W_A$   &\, $W_B$                  & \quad  & phase   \\

\hline

$|+\rangle_a$  &  $|\mu(\theta)\rangle_b$   & \quad &   $R_z(\alpha+\frac{\pi}{2})$     & \, $R_\mathbf{n}(\frac{\pi}{2}-\theta)$                  &        & $e^{\texttt{i}\frac{\alpha}{2}}$ \\

$|+\rangle_a$  &  $|\nu(\theta)\rangle_b$   & \quad &   $R_z(\alpha-\frac{\pi}{2})$     & \, $R_\mathbf{n}(-\frac{\pi}{2}-\theta)$                  &        & $\texttt{i}e^{\texttt{i}\frac{\alpha}{2}}$ \\

$|-\rangle_a$  &  $|\nu(\pi-\theta)\rangle_b$   & \quad &   $R_z(\alpha+\frac{\pi}{2})$     & \, $R_\mathbf{n}(\frac{\pi}{2}-\theta)$                 &        & $e^{\texttt{i}\frac{\alpha}{2}}$\\

$|-\rangle_a$  &  $|\mu(\pi-\theta)\rangle_b$   & \quad &  $R_z(\alpha-\frac{\pi}{2})$     & \,  $R_\mathbf{n}(-\frac{\pi}{2}-\theta)$                    &        & $\texttt{i}e^{\texttt{i}\frac{\alpha}{2}}$ \\
                             \hline
\end{tabular}\label{table1}
\end{table}

Putting all the pieces together, one can see that the protocol shown in figure \ref{nonlocalCU} completes an arbitrary nonlocal CU gate teleportation in a deterministic way. This protocol requires one pre-shared ebit and two cbits, and employs a superposition of single-qubit gate orders, achieving the minimal communication resource cost \cite{jiang2007distributed,eisert2000optimal}.
Unlike the previous CU gate teleportation in references~\cite{eisert2000optimal,liu2024nonlocal}, which involve experimentally challenging local CNOT and CU operations at each node, the proposed protocol avoids these complexities by exploiting single-qubit operations via the ICO framework.
Specifically,
when the parameters are set as
     $\alpha=-\frac{\pi}{2}$, $\theta=\frac{\pi}{2}$, $\mathbf{n}=(1, 0, 0)$, and $\mathbf{n}^\perp=(0, 0, 1)$, the protocol implements CNOT gate teleportation.
When $\alpha=-\frac{\pi}{2}$, $\theta=\frac{\pi}{2}$, $\mathbf{n}=(0, 0, 1)$, and $\mathbf{n}^\perp=(1, 0, 0)$, the protocol implements controlled-$Z$ gate teleportation.
When $\alpha=-\frac{\pi}{2}$, $\theta=\frac{\pi}{2}$, $\mathbf{n}=(0, 1, 0)$, and $\mathbf{n}^\perp=(1, 0, 0)$, the protocol implements controlled-$Y$ gate teleportation.
When $\alpha=-\frac{\pi}{2}$, $\theta=\frac{\pi}{2}$, $\mathbf{n}=\frac{1}{\sqrt{2}}(1, 0, 1)$, and $\mathbf{n}^\perp=(0, 1, 0)$, the protocol implements controlled-$H$ gate teleportation.
Here $Z$ and $Y$ denote Pauli $Z$ and  Pauli $Y$ operators, respectively, and $H$ denotes the Hadamard gate. It is worth noting that the choice of $\mathbf{n}^\perp$ is not unique in general.
When multiple CU gates are teleported, the accumulated overall phase from each CU gate remains trivial and does not affect the arguments.

\section{The optical realization of an arbitrary CU gate teleportation} \label{Sec3}

\begin{figure} 
\begin{center}
\includegraphics[width=10.2 cm,angle=0]{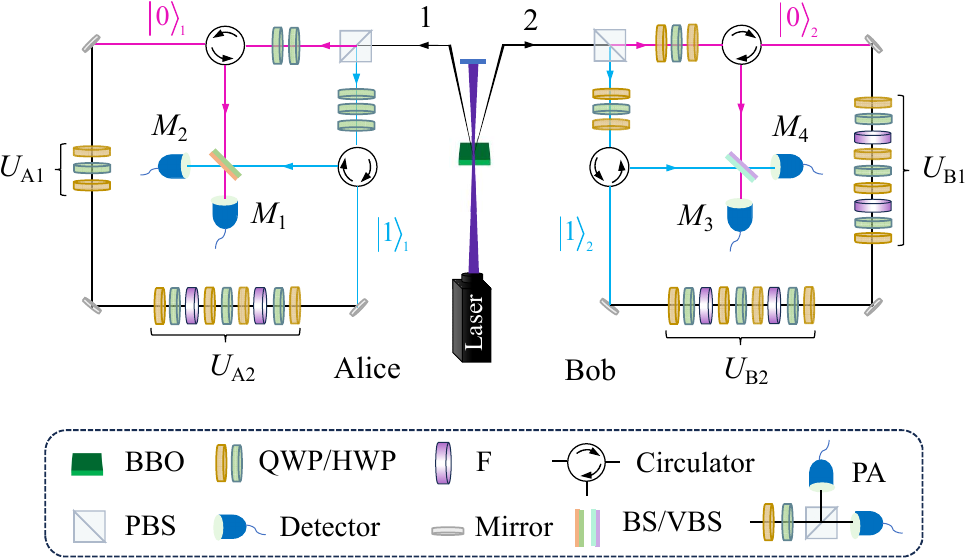}
\caption{\justifying
Optical setup for teleporting an arbitrary two-qubit CU gate on the polarization DOF. A polarization-entangled photon state $|\Psi\rangle_{0}=\frac{1}{\sqrt{2}}(|H\rangle_1|H\rangle_2+\texttt{i}|V\rangle_1|V\rangle_2)$ is generated via a type-I spontaneous parametric down-conversion (SPDC) process. The photon pairs then pass through polarization beam splitters (PBSs), half-wave plates (HWPs), and quarter-wave plates (QWPs) to convert the polarization-entangled state into a general path-entangled state $|\Psi\rangle_{3}=\frac{1}{\sqrt{2}}(|0\rangle_1|0\rangle_2+\texttt{i}|1\rangle_1|1\rangle_2)|\psi\rangle_1'|\psi\rangle_2'$, by exploiting the fact that the PBS transmits $H$-polarized photons while reflecting $V$-polarized photons. The photons are subsequently directed into a reciprocal polarization Sagnac interferometer to implement a superposition of the orders of the single-qubit polarization gates $U_{A_1}$, $U_{A_2}$, $U_{B_1}$, and $U_{B_2}$. Finally, photons from different paths are recombined at either a balanced beam splitter (BS) or a variable beam splitter (VBS). Coincidence measurements on the path states $|0\rangle_1|0\rangle_2$, $|1\rangle_1|1\rangle_2$, $|0\rangle_1|1\rangle_2$, and $|1\rangle_1|0\rangle_2$ are performed by using the detector pairs $\{M_1, M_3\}$, $\{M_2, M_4\}$, $\{M_1, M_4\}$, and $\{M_2, M_3\}$, respectively. Here, each $M_i$ comprises a polarization analyzer (PA) capable of reconstructing the polarization state in any basis.
} \label{setup}
\end{center}
\end{figure}

Quantum switches are the powerful tools for achieving ICO.
However, most existing implementations rely on folded Mach-Zehnder interferometers (MZIs) \cite{procopio2015experimental}, which are prone to phase instability and polarization-dependent effects due to the presence of non-reciprocal optical elements.
These limitations can significantly degrade the fidelity and efficiency of the resulting operations.
To address these challenges, we employ a fully reciprocal Sagnac-loop interferometer \cite{liu2023experimentally,stromberg2023demonstration} to implement the teleportation of a CU gate.

Figure \ref{setup} shows the setup for implementing CU gate teleportation using a common-path Sagnac interferometric configuration.
In this architecture, the computational qubits are encoded in the polarization DOF of single photons.
The coherent superposition of single-qubit polarization gate operations is realized by routing photons along clockwise and counterclockwise paths within the interferometer.

The setup begins with a 405 nm continuous-wave laser that pumps a $\beta$-barium borate (BBO, $\beta$-BaB$_2$O$4$) crystal to generate polarization-entangled photon pairs via a type-I spontaneous parametric down-conversion (SPDC) process \cite{zhang2021spontaneous}.
The resulting two-photon state is characterized by
\begin{eqnarray}
\begin{split}             \label{eq14}
|\Psi\rangle_{0}=\frac{1}{\sqrt{2}}(|H\rangle_1|H\rangle_2+\texttt{i}|V\rangle_1|V\rangle_2),
\end{split}
\end{eqnarray}
where $H$ and $V$ denote the horizontal and vertical polarizations, respectively. Subscripts 1 and 2 label the spatial modes (paths) of the two photons.

As illustrated in figure \ref{setup}, the photon in path 1 is directed to Alice, while that in path 2 is sent to Bob.
In each laboratory, the photons first pass through polarization beam splitters (PBSs), which transmit the $H$-polarized components into the path states $|0\rangle_1$ and $|0\rangle_2$ (indicated by the pink circuits), and reflect the $V$-polarized components into the path states $|1\rangle_1$ and $|1\rangle_2$ (indicated by the blue circuits). After this process, the entangled state $|\Psi\rangle_0$ is transformed into
\begin{eqnarray}
\begin{split}             \label{eq15}
|\Psi\rangle_{1}=\frac{1}{\sqrt{2}}(|H\rangle|0\rangle_1|H\rangle|0\rangle_2+\texttt{i}|V\rangle|1\rangle_1|V\rangle|1\rangle_2).
\end{split}
\end{eqnarray}

Second, to create the desired initial state,  two half-wave plates (HWPs) oriented at $45^\circ$ are placed in the paths $|1\rangle_1$ and $|1\rangle_2$ to flip the $V$-polarized photons into $H$-polarized ones, i.e., $|V\rangle|1\rangle_1 \rightarrow |H\rangle|1\rangle_1$ and $|V\rangle|1\rangle_2 \rightarrow |H\rangle|1\rangle_2$.
Following this, HWPs oriented at $\theta_1^\circ$ are placed in both paths $|0\rangle_1$ and $|1\rangle_1$ at Alice's site, while HWPs oriented at $\theta_2^\circ$ are placed in both paths $|0\rangle_2$ and $|1\rangle_2$ at Bob's site, to prepare arbitrary single-qubit polarization states.
These operations transform $|\Psi\rangle_{1}$ into
\begin{eqnarray}
\begin{split}             \label{eq16}
|\Psi\rangle_{2}=\frac{1}{\sqrt{2}}(|0\rangle_1|0\rangle_2+\texttt{i}|1\rangle_1|1\rangle_2)|\psi\rangle_1|\psi\rangle_2.
\end{split}
\end{eqnarray}
Here $|\psi\rangle_1$ and $|\psi\rangle_2$ are general single-qubit polarization states, and they are given by
\begin{eqnarray}
\begin{split}             \label{eq17}
|\psi\rangle_1 =\cos\theta_1|H\rangle_1+\sin\theta_1|V\rangle_1,
\end{split}
\end{eqnarray}
\begin{eqnarray}
\begin{split}             \label{eq18}
|\psi\rangle_2 =\cos\theta_2|H\rangle_2+\sin\theta_2|V\rangle_2.
\end{split}
\end{eqnarray}

Third, single-qubit polarization gates are applied.
On Alice's side, the $V_A$ gates are realized by placing HWPs oriented at $45^\circ$ in both paths $|0\rangle_1$ and $|1\rangle_1$.
On Bob's side, the $V_B$ gates, along with the transformations in the second step, are jointly realized by inserting two quarter-wave plates (QWPs) and one HWP into each of the paths $|0\rangle_2$ and $|1\rangle_2$.
These optical elements transform the state $|\Psi\rangle_{2}$ into
\begin{eqnarray}
\begin{split}             \label{eq19}
|\Psi\rangle_{3}=\frac{1}{\sqrt{2}}(|0\rangle_1|0\rangle_2+\texttt{i}|1\rangle_1|1\rangle_2)|\psi\rangle_1'|\psi\rangle_2'.
\end{split}
\end{eqnarray}
Here $|\psi\rangle_1'=V_A|\psi\rangle_1$ and $|\psi\rangle_2'=V_B|\psi\rangle_2$.
The synthesis of HWP and QWP in term of single-qubit rotations can be expressed as  \cite{simon1990minimal}
\begin{eqnarray}
\begin{split}             \label{eq21}
\text{HWP}(\theta) = R_y(2\theta) \cdot R_z(\pi)   \cdot R_y(-2\theta),
\end{split}
\end{eqnarray}
\begin{eqnarray}
\begin{split}             \label{eq20}
\text{QWP}(\theta) = R_y(2\theta) \cdot R_z(\pi/2) \cdot R_y(-2\theta).
\end{split}
\end{eqnarray}


Fourth, a stable Sagnac interferometer with perfect path overlap is used to implement the quantum switch.
In the architecture, the path DOF acts as a control qubit that governs  the order of the single-qubit polarization gate operations.
On Alice's side, the photon in path $|0\rangle_1$ travels counterclockwise, encountering the $U_{A_1}$ gate first followed by $U_{A_2}$ (i.e., $U_{A_2}U_{A_1}$),
whereas the photon in path $|1\rangle_1$ travels clockwise, undergoing $U_{A_2}$ before $U_{A_1}$ (i.e., $U_{A_1}U_{A_2}$).
The two paths are then recombined on a balanced beam splitter (BS).
A similar arrangement is made on Bob's side, whereas the paths are fed into a variable beam splitter (VBS).
Therefore, the operations $U_{A_1}$, $U_{A_2}$, $U_{B_1}$, $U_{B_2}$, BS, and VBS convert $|\Psi\rangle_{3}$ to
\begin{eqnarray}
\begin{split}             \label{eq22}
|\Psi\rangle_{4}=&\frac{1}{2}\big\{|0\rangle_1|0\rangle_2[\cos(\frac{\theta}{2})(U_{A_2} U_{A_1})\otimes (U_{B_2} U_{B_1})
                                               +\texttt{i}\sin(\frac{\theta}{2})(U_{A_1} U_{A_2})\otimes (U_{B_1} U_{B_2})]|\psi\rangle_1'|\psi\rangle_2'\\
                 &+|0\rangle_1|1\rangle_2[\sin(\frac{\theta}{2})(U_{A_2} U_{A_1})\otimes (U_{B_2} U_{B_1})
                               -\texttt{i}\cos(\frac{\theta}{2})(U_{A_1} U_{A_2})\otimes (U_{B_1} U_{B_2})]|\psi\rangle_1'|\psi\rangle_2'\\
                 &+|1\rangle_1|0\rangle_2[\cos(\frac{\theta}{2})(U_{A_2} U_{A_1})\otimes (U_{B_2} U_{B_1})
                               -\texttt{i}\sin(\frac{\theta}{2})(U_{A_1} U_{A_2})\otimes (U_{B_1} U_{B_2})]|\psi\rangle_1'|\psi\rangle_2'\\
                 &+|1\rangle_1|1\rangle_2[\sin(\frac{\theta}{2})(U_{A_2} U_{A_1})\otimes (U_{B_2} U_{B_1})
                               +\texttt{i}\cos(\frac{\theta}{2})(U_{A_1} U_{A_2})\otimes (U_{B_1} U_{B_2})]|\psi\rangle_1'|\psi\rangle_2'\big\}.
\end{split}
\end{eqnarray}
Here the transformations of the polarization-independent VBS is given by
%
\begin{eqnarray}
\begin{split}             \label{eq23}
|0\rangle_2 \xrightarrow{\text{VBS}}\cos(\frac{\theta}{2})|0\rangle_2 + \sin(\frac{\theta}{2})|1\rangle_2,  \quad
|1\rangle_2 \xrightarrow{\text{VBS}}\sin(\frac{\theta}{2})|0\rangle_2 - \cos(\frac{\theta}{2})|1\rangle_2.
\end{split}
\end{eqnarray}
%
The VBS can be simulated by two 50:50 BS and two phase shifters. The BS is a special VBS with equal transmittance and reflectance of $1/\sqrt{2}$.

It is noteworthy that the Sagnac-based quantum switch is implemented using a common-path configuration combined with a fully reciprocal polarization gadget. In this setup, due to the common-path propagation of photons, the relative mode phase is very small, which is highly robust against phase drifts. Moreover, in principle, the forward- and backward-propagating photons undergo identical polarization transformations.
Importantly, such reciprocity cannot be generally achieved using conventional QWP–HWP–QWP sequences.
The above reciprocal polarization gadget, capable of simulating an arbitrary unitary operation $U \in \text{SU}(2)$, can be constructed as \cite{stromberg2023demonstration}
\begin{eqnarray}
\begin{split}             \label{eq24}
U = \; &  \text{QWP}(\theta_1)
    \cdot \text{HWP}(\phi_1)
    \cdot R_y(-\frac{\pi}{2})
    \cdot \text{QWP}(\frac{\pi}{2})\\&
    \cdot \text{HWP}(\gamma)
    \cdot \text{QWP}(\frac{\pi}{2})
    \cdot  R_y(\frac{\pi}{2})
    \cdot \text{HWP}(\phi_2)
    \cdot \text{QWP}(\theta_2).
\end{split}
\end{eqnarray}
For example, the reciprocal optical implementation of the gate
       $U_{A_1} = R_z(\frac{\pi}{2})= \text{QWP}(\frac{\pi}{4}) \cdot \text{HWP}(\frac{3\pi}{8}) \cdot \text{QWP}(\frac{\pi}{4})$,
while the implementation of the
$U_{A_2}= X =  \text{QWP}(\frac{3\pi}{4})
    \cdot \text{HWP}(\frac{7\pi}{8})
    \cdot   R_y(-\frac{\pi}{2})
    \cdot \text{QWP}(\frac{\pi}{2})
    \cdot \text{HWP}(\frac{3\pi}{4})
    \cdot \text{QWP}(\frac{\pi}{2})
    \cdot   R_y(\frac{\pi}{2})
    \cdot \text{HWP}(\frac{3\pi}{8})
    \cdot \text{QWP}(\frac{\pi}{4})$.
The rotation operators $R_y\left(\pm\frac{\pi}{2}\right)$ can be physically implemented using Faraday rotators with a circular retardance of $\pm\frac{\pi}{2}$.


Finally, the coincidence measurements are performed on the path DOF of the output state $|\Psi\rangle_{4}$.
Specifically, the two-fold coincidence counting events, i.e., $\{M_1, M_3\}$, $\{M_2, M_4\}$, $\{M_1, M_4\}$, and $\{M_2, M_3\}$, correspond to path states $|0\rangle_1|0\rangle_2$, $|1\rangle_1|1\rangle_2$, $|0\rangle_1|1\rangle_2$, and $|1\rangle_1|0\rangle_2$, respectively.
The parameters of the VBS are adaptively adjusted based on Alice's measurement outcomes.
In detail,
if detector $M_1$ (associated with path $|0\rangle_1$) clicks, the VBS is set with transmittance $\cos(\theta/2)$ and reflectance $\sin(\theta/2)$; if $M_2$ (associated with path $|1\rangle_1$) clicks, the settings are $\sin(\theta/2)$ and $\cos(\theta/2)$, respectively.
Each path measurement $M_i$ ($i=1,\ldots,4$)  is accompanied by polarization state tomography using a polarization analyzer (PA) composed of one QWP, one HWP, one PBS, and two single-photon detectors. This enables complete reconstruction of the output polarization state.

\section{Discussion and conclusion} \label{Sec4}

\begin{figure} 
\begin{center}
\includegraphics[width=8.7 cm,angle=0]{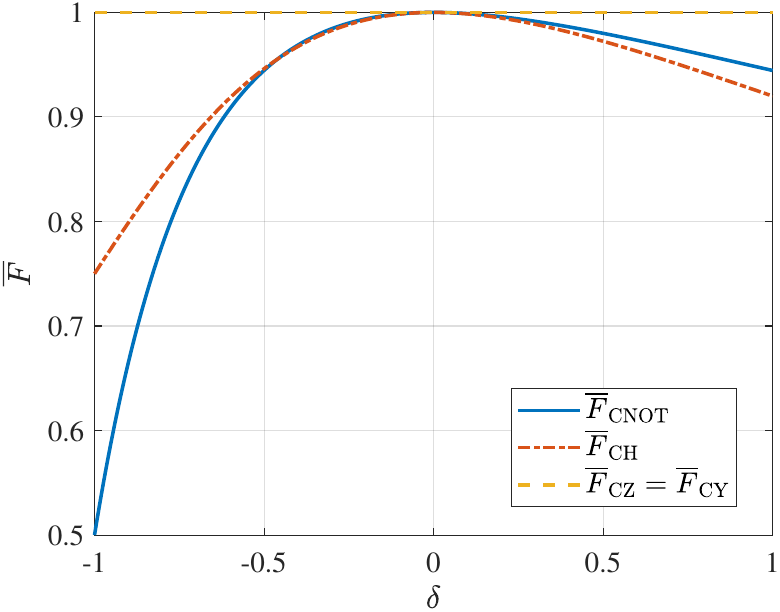}
\caption{\justifying
Relations between the average gate fidelity $\overline{F}$ of various CU gates and the  imperfection parameter $\delta$. When $\delta=0$, all gate fidelities are unity. In addition,
 $\overline{F}_\text{CY}=\overline{F}_\text{CZ}=1$ for any value of $\delta$.} \label{fidelity}
\end{center}
\end{figure}

In our scheme, a reciprocal optical implementation of the gate can be achieved in principle. However, in practical experiments, imperfect reciprocity may arise during forward and backward photon propagation. Such imperfections can be characterized as deviations from the ideal transformation and described by the mapping $[\theta, n_x, n_y, n_z]\mapsto[\theta, (1+\delta)n_x, n_y, n_z]$ for each single-qubit gate \cite{stromberg2023demonstration}. Under imperfect reciprocity (i.e., the imperfection parameter $\delta\neq0$), the practical implementations of the gates $U_{A_1}$, $U_{A_2}$, $U_{B_1}$, and $U_{B_2}$ are given by
\begin{eqnarray}
\begin{split}             \label{eq27}
\tilde{U}_{A_1} =\left(\begin{array}{cc}
          e^{-\texttt{i}\frac{\pi}{4}}    &           0                 \\
                 0                        &  e^{\texttt{i}\frac{\pi}{4}}\\
\end{array}\right),\\
\end{split}
\end{eqnarray}
\begin{eqnarray}
\begin{split}             \label{eq28}
\tilde{U}_{A_2}=(1+\delta)\left(\begin{array}{cc}
           0  &  1  \\
           1  &  0  \\
\end{array}\right), \\
\end{split}
\end{eqnarray}
\begin{eqnarray}
\begin{split}             \label{eq29}
\tilde{U}_{B_1}=\frac{1}{\sqrt{2}} \left(\begin{array}{cc}
             1-\texttt{i} n_z            &  -\texttt{i}(1+\delta) n_x-n_y\\
            -\texttt{i}(1+\delta) n_x+n_y  &   1+\texttt{i} n_z     \\
\end{array}\right),\\
\end{split}
\end{eqnarray}
and
\begin{eqnarray}
\begin{split}             \label{eq30}
\tilde{U}_{B_2}=\left(\begin{array}{cc}
           n_z'  &  (1+\delta)n_x'-\texttt{i} n_y'  \\
           (1+\delta)n_x'+\texttt{i} n_y'  &  -n_z'  \\
\end{array}\right).
\end{split}
\end{eqnarray}
To evaluate the performance of the CU gate, we calculate its average fidelity, defined as
\begin{eqnarray}
\begin{split}             \label{eq225}
\overline{F}_\text{CU}=\frac{1}{(2\pi)^2}\int_{0}^{2\pi}\int_{0}^{2\pi}|\langle\Psi_\text{ideal}|\Psi_\text{prac}\rangle|^2d\theta_1d\theta_2.
\end{split}
\end{eqnarray}
Here $|\Psi_{\mathrm{ideal}}\rangle$ denotes the ideal photonic output state using the ideal single-qubit gates $U_{A_1}$, $U_{A_2}$, $U_{B_1}$, and $U_{B_2}$, as defined in equations~\eqref{eq11}-\eqref{eq11.3}, while $|\Psi_{\mathrm{prac}}\rangle$ denotes the corresponding practical output state using the imperfect gates $\tilde{U}_{A_1}$, $\tilde{U}_{A_2}$, $\tilde{U}_{B_1}$, and $\tilde{U}_{B_2}$ given in equations~\eqref{eq27}–\eqref{eq30}. The average fidelities of the CNOT, CY, CZ, and CH gates are plotted in figure~\ref{fidelity}. In the ideal case, i.e., for $\delta = 0$, all average gate fidelities are unity. Moreover, $\overline{F}_{\mathrm{CY}} = \overline{F}_{\mathrm{CZ}} = 1$ for any value of $\delta$, indicating that these two gates are unaffected by the imperfection parameter.


In conclusion, we have presented an efficient protocol for teleporting arbitrary CU gates between two spatially separated nodes. Our protocol is accomplished by exploiting ICOs, one ebit, and two cbits as the resource, which achieving the minimal communication cost \cite{jiang2007distributed,eisert2000optimal}.
In conventional fixed-gate-order approaches, the teleportation of a CU gate typically requires one local CNOT gate and one local CU gate \cite{eisert2000optimal,liu2024nonlocal}. The implementation of the CU gate is more complex, as it has to be decomposed into two CNOT gates and three fundamental single-qubit gates \cite{barenco1995elementary}. If the based-ICO CNOT gate teleportation \cite{liu2025quantum} is directly combined with this decomposition, the resulting nonlocal CU gate would consume two shared ebits, four cbits, and four quantum switches, thereby  doubling the resource cost compared with our approach.


In contrast, our protocol is implemented by appropriately adjusting the inherent single-qubit gates. This significantly reduces the resource overhead and offers improved flexibility compared to fixed causal-order circuits and conventional gate-decomposition strategies.
Furthermore, we proposed an optical implementation of CU gate teleportation based on a common-path Sagnac loop interferometer.
This Sagnac-based design inherently suppresses phase instabilities and polarization-dependent effects through reciprocal photon propagation.
Since different qubit-like DOFs are used for encoding, the photons are ultimately detected to verify the CU gate, which may limit scalability. This issue may be solved by directly routing the photons to subsequent gate operations without measurement, or by employing nondestructive photonic qubit detection techniques \cite{Nondestructive1,Nondestructive2}. Furthermore, the protocol is not restricted to photonic systems and can be extended to other qubit candidates, such as neutral atoms \cite{daiss2021quantum} and trapped ions \cite{main2025distributed}, where nondestructive quantum operations and broader applications are feasible.

%
%


\data{The data that support the findings of this study are available upon reasonable request from the authors.}

\ack{This work was supported by the National Natural Science Foundation of China under Grant No. 12505028 and Grant No. 62371038, and  Science Research Project of Hebei Education Department under Grant No. QN2025054.}

\appendix
\setcounter{equation}{0}
\renewcommand{\theequation}{A\arabic{equation}}

\section*{Appendix Proof of nonlocal CU gate implementation}  \label{Appendix}

In section \ref{Sec2}, we have shown that the protocol illustrated in figure \ref{nonlocalCU} enables the teleportation of an arbitrary two-qubit CU gate.

Single-qubit gates
\begin{eqnarray}
\begin{split}             \label{A}
&V_{A}\otimes V_{B}=X\otimes(\mathbf{n}^\perp\cdot \bm{\sigma}),\\
&W_{A}^{\mu}\otimes W_{B}^{\mu}=R_z(\alpha+\frac{\pi}{2})\otimes R_\mathbf{n}(\frac{\pi}{2}-\theta),\\
&W_{A}^{\nu}\otimes W_{B}^{\nu}=R_z(\alpha-\frac{\pi}{2})\otimes R_\mathbf{n}(-\frac{\pi}{2}-\theta),
\end{split}
\end{eqnarray}
and the superposition operations of single-qubit gates $S_{+,\mu}^{A,B}(\theta)$ and $S_{+,\nu}^{A,B}(\theta)$
are employed to construct our CU gate via quantum switches.
The operations $S_{+,\mu}^{A,B}(\theta)$ and $S_{+,\nu}^{A,B}(\theta)$  in equations \eqref{eq9} and \eqref{eq10}, are given by
\begin{eqnarray}
\begin{split}             \label{A1}
S_{+,\mu}^{A,B}(\theta)=\cos\big(\frac{\theta}{2}\big) U_{A_2}U_{A_1}\otimes U_{B_2}U_{B_1}  +\texttt{i} \sin\big(\frac{\theta}{2}\big)  U_{A_1}U_{A_2}\otimes U_{B_1}U_{B_2},
\end{split}
\end{eqnarray}
and
\begin{eqnarray}
\begin{split}             \label{A2}
S_{+,\nu}^{A,B}(\theta)=\sin\big(\frac{\theta}{2}\big) U_{A_2}U_{A_1}\otimes U_{B_2}U_{B_1}   -\texttt{i} \cos\big(\frac{\theta}{2}\big)  U_{A_1}U_{A_2}\otimes U_{B_1}U_{B_2}.
\end{split}
\end{eqnarray}
Next, we prove that for the different measured outcomes on qubits $a$ and $b$, the arbitrary CU gate can be equivalently constructed as
\begin{equation}  \label{A3}
\text{CU}=\left\{\begin{aligned}
&e^{\texttt{i}\frac{\alpha}{2}}(W_{A}^{\mu}\otimes W_{B}^{\mu})S_{+,\mu}^{A,B}(\theta)(V_{A}\otimes V_{B}),  \\
&\texttt{i}e^{\texttt{i}\frac{\alpha}{2}}(W_{A}^{\nu}\otimes W_{B}^{\nu})S_{+,\nu}^{A,B}(\theta)(V_{A}\otimes V_{B}). \\
\end{aligned}
\right.
\end{equation}

\begin{proof} An arbitrary two-qubit CU gate is defined as \cite{nielsen2010quantum}
\begin{eqnarray}  \label{A4}
\text{CU} = |0\rangle_A\langle0|\otimes \mathbb{I}_B+|1\rangle_A\langle1|\otimes U_B(\alpha, \theta, \textbf{n}),
\end{eqnarray}
where $U_B(\alpha, \theta, \textbf{n})$ is an arbitrary single-qubit unitary that acts on the target qubit $B$ when control qubit $A$ is in the state $|1\rangle$, and the target qubit remains unchanged when $A$ is in the state $|0\rangle$.
Here $\alpha$, $\theta \in \mathbb{R}$ are real parameters, $\textbf{n}=(n_x, n_y, n_z)\in \mathbb{R}^3$ is a real unit vector, and $\bm{\sigma}=(X, Y, Z)$ represents the Pauli vector. The single-qubit gate $U_B(\alpha, \theta, \textbf{n})$ is given by
\begin{eqnarray}
\begin{split}             \label{A5}
U_B(\alpha, \theta, \textbf{n})=\text{exp}\big[\texttt{i}\big(\alpha \mathbb{I}+\theta(\textbf{n}\cdot \bm{\sigma})\big)\big].
\end{split}
\end{eqnarray}
The CU gate can be equivalently rewritten in exponential form and decomposed as follows
\begin{eqnarray}
\begin{split}             \label{A6}
\text{CU}=\text{exp}\big[\texttt{i}|1\rangle\langle1|\otimes \big(\alpha \mathbb{I}+\theta(\textbf{n}\cdot \bm{\sigma})\big)\big]
=e^{i\frac{\alpha}{2}}\big(R_z(\alpha)\otimes R_\mathbf{n}(-\theta)\big)R_{z\mathbf{n}}(\theta).
\end{split}
\end{eqnarray}
Here $R_\mathbf{n}(\theta)$ denotes a single-qubit rotation by an angle $\theta$ around the \textbf{n}-axis, as defined in equation \eqref{eq13}.
Similarly, $R_z(\theta)$ is a rotation around the vector $\mathbf{n}_z=(0, 0, 1)$ i.e., $z$-axis.
$R_{z\mathbf{n}}(\theta)$  represents a two-qubit rotation gate by an angle $\theta$, simultaneously involving the around the vectors $\mathbf{n}_z$ and \textbf{n}.
The two-qubit rotation $R_{z\mathbf{n}}(\theta)$ is expressed as
\begin{eqnarray}
\begin{split}             \label{A7}
R_{z\mathbf{n}}(\theta)=\cos\big(\frac{\theta}{2}\big)\mathbb{I}\otimes\mathbb{I} -\texttt{i}\sin\big(\frac{\theta}{2}\big)(\mathbf{n}_z\cdot \bm{\sigma})\otimes (\mathbf{n}\cdot \bm{\sigma}).
\end{split}
\end{eqnarray}
We define the following single-qubit gates for the decomposition
\begin{eqnarray}
\begin{split}             \label{A8}
&V_A=X, \qquad  V_B=\mathbf{n}^\perp\cdot \bm{\sigma}, \qquad  U_{A_1}=R_z\big(\frac{\pi}{2}\big),\\&
U_{A_2}=X, \quad\;\; U_{B_2}=\mathbf{n}^\perp\cdot \bm{\sigma}, \quad\;\; U_{B_1}=R_\mathbf{n}\big(\frac{\pi}{2}\big).
\end{split}
\end{eqnarray}
From these above definitions, the following identities can be derived
\begin{eqnarray}
\begin{split}             \label{A9}
&U_{A_2}U_{A_1}V_A=R_z(-\frac{\pi}{2}), \quad\; U_{A_1}U_{A_2}V_A=R_z(\frac{\pi}{2}), \\
&U_{B_2}U_{B_1}V_B=R_\mathbf{n}(-\frac{\pi}{2}),  \quad\; U_{B_1}U_{B_2}V_B=R_\mathbf{n}(\frac{\pi}{2}).
\end{split}
\end{eqnarray}
Substituting  equation \eqref{A9} into equation \eqref{A3}, we obtain
\begin{eqnarray}
\begin{split}             \label{A10}
S_{+,\mu}^{A,B}(\theta)(V_{A}\otimes V_{B})=& \cos\big(\frac{\theta}{2}\big) R_z(-\frac{\pi}{2})\otimes  R_\mathbf{n}(-\frac{\pi}{2})   +\texttt{i} \sin\big(\frac{\theta}{2}\big)  R_z(\frac{\pi}{2})\otimes  R_\mathbf{n}(\frac{\pi}{2}),
\end{split}
\end{eqnarray}
and
\begin{eqnarray}
\begin{split}             \label{A11}
S_{+,\nu}^{A,B}(\theta)(V_{A}\otimes V_{B})=& \sin\big(\frac{\theta}{2}\big) R_z(-\frac{\pi}{2})\otimes  R_\mathbf{n}(-\frac{\pi}{2})   -\texttt{i} \cos\big(\frac{\theta}{2}\big)  R_z(\frac{\pi}{2})\otimes  R_\mathbf{n}(\frac{\pi}{2}).
\end{split}
\end{eqnarray}
We further introduce the following single-qubit gates
\begin{eqnarray}
\begin{split}             \label{A12}
&\tilde{W}_{A}^{\mu}=R_z(\frac{\pi}{2}), \quad \tilde{W}_{B}^{\mu}=R_\mathbf{n}(\frac{\pi}{2}),  \quad
\tilde{W}_{A}^{\nu}=\texttt{i}R_z(-\frac{\pi}{2}), \quad \tilde{W}_{B}^{\nu}=R_\mathbf{n}(-\frac{\pi}{2}),
\end{split}
\end{eqnarray}
to construct the operator $R_{z\mathbf{n}}(\theta)$. Based on the identities $R_\mathbf{n}(\pm\frac{\pi}{2})R_\mathbf{n}(\mp\frac{\pi}{2})=\mathbb{I}$, $R_\mathbf{n}(\pm\frac{\pi}{2})R_\mathbf{n}(\pm\frac{\pi}{2})=R_\mathbf{n}(\pm\pi)$, and
$R_z(\pm\pi)\otimes R_\mathbf{n}(\pm\pi)=-Z\otimes(\mathbf{n}\cdot\sigma)$, it is easily to verify
\begin{equation}  \label{A13}
R_{z\mathbf{n}}(\theta)=\left\{\begin{aligned}
&(\tilde{W}_{A}^{\mu}\otimes \tilde{W}_{B}^{\mu})S_{+,\mu}^{A,B}(\theta)(V_{A}\otimes V_{B}),  \\
&(\tilde{W}_{A}^{\nu}\otimes \tilde{W}_{B}^{\nu})S_{+,\nu}^{A,B}(\theta)(V_{A}\otimes V_{B}).
\end{aligned}
\right.
\end{equation}
Finally, combining with equation \eqref{A6}, the CU gate can be fully expressed as
\begin{equation}  \label{A14}
\text{CU}=\left\{\begin{aligned}
&e^{\texttt{i}\frac{\alpha}{2}}(W_{A}^{\mu}\otimes W_{B}^{\mu}) S_{+,\mu}^{A,B}(\theta)(V_{A}\otimes V_{B}),  \\
&\texttt{i}e^{\texttt{i}\frac{\alpha}{2}}(W_{A}^{\nu}\otimes W_{B}^{\nu}) S_{+,\mu}^{A,B}(\theta)(V_{A}\otimes V_{B}). \\
\end{aligned}
\right.
\end{equation}
Here $W_{A}^{\mu}=R_z(\alpha+\frac{\pi}{2})$, $W_{B}^{\mu}=R_\mathbf{n}(\frac{\pi}{2}-\theta)$, $W_{A}^{\nu}=R_z(\alpha-\frac{\pi}{2})$, and $W_{B}^{\nu}=R_\mathbf{n}(-\frac{\pi}{2}-\theta)$. \qedhere
\end{proof}







\providecommand{\newblock}{}

\end{document}